\documentclass[10pt,twocolumn,superscriptaddress,english,10pt,showpacs,floatfix,prl]{revtex4-1}
\usepackage{amsthm}
\usepackage{amsmath}
\usepackage{amssymb}
\usepackage{mathdots}
\usepackage{graphicx}
\usepackage[normalem]{ulem}
\usepackage{array}
\makeatletter

\@ifundefined{textcolor}{}
{%
 \definecolor{BLACK}{gray}{0}
 \definecolor{WHITE}{gray}{1}
 \definecolor{RED}{rgb}{1,0,0}
 \definecolor{GREEN}{rgb}{0,1,0}
 \definecolor{BLUE}{rgb}{0,0,1}
 \definecolor{CYAN}{cmyk}{1,0,0,0}
 \definecolor{MAGENTA}{cmyk}{0,1,0,0}
 \definecolor{YELLOW}{cmyk}{0,0,1,0}
}

\usepackage{braket}
\usepackage[backref=true,
bookmarksnumbered=true,
bookmarks=true,
bookmarksopen=true,
colorlinks=true,
citecolor=blue,
linkcolor=blue,
anchorcolor=green,
urlcolor=blue,unicode=false]{hyperref}
\newcolumntype{L}[1]{>{\raggedright\let\newline\\\arraybackslash\hspace{0pt}}m{#1}}
\newcolumntype{C}[1]{>{\centering\let\newline\\\arraybackslash\hspace{0pt}}m{#1}}
\newcolumntype{R}[1]{>{\raggedleft\let\newline\\\arraybackslash\hspace{0pt}}m{#1}}

\newcommand{\abs}[1]{\left| #1 \right|} 
 
\let\baraccent=\= 
\renewcommand{\=}[1]{\stackrel{#1}{=}} 

\DeclareMathOperator\Sgn{Sgn}

\makeatother

\usepackage{babel}

\begin{document}
\title{Parity anomaly and spin transmutation in quantum spin Hall Josephson junctions}

\author{Yang Peng}
\affiliation{\mbox{Dahlem Center for Complex Quantum Systems and Fachbereich Physik, Freie Universit{\"a}t Berlin, 14195 Berlin, Germany} }

\author{Yuval Vinkler-Aviv}
\affiliation{\mbox{Dahlem Center for Complex Quantum Systems and Fachbereich Physik, Freie Universit{\"a}t Berlin, 14195 Berlin, Germany} }

\author{Piet W.\ Brouwer}
\affiliation{\mbox{Dahlem Center for Complex Quantum Systems and Fachbereich Physik, Freie Universit{\"a}t Berlin, 14195 Berlin, Germany} }

\author{Leonid I.\ Glazman}
\affiliation{Department of Physics, Yale University, New Haven, Connecticut 06520, USA}

\author{Felix von Oppen}
\affiliation{\mbox{Dahlem Center for Complex Quantum Systems and Fachbereich Physik, Freie Universit{\"a}t Berlin, 14195 Berlin, Germany} }

\begin{abstract}
We study the Josephson effect in a quantum spin Hall system coupled to a localized magnetic impurity. As a consequence of the fermion parity anomaly, the spin of the combined system of impurity and spin-Hall edge alternates between half-integer and integer values when the superconducting phase difference across the junction advances by $2\pi$. This leads to characteristic differences in the splittings of the spin multiplets by exchange coupling and single-ion anisotropy at phase differences, for which time-reserval symmetry is preserved.  We discuss the resulting $8\pi$-periodic (or $\mathbb{Z}_4$) fractional Josephson effect in the context of recent experiments. \end{abstract}

\maketitle

{\em Introduction.---}The fractional Josephson effect \cite{kitaev01, kwon04, fu09} constitutes one of the most striking effects heralding topological superconductivity \cite{review1,review2}. In Josephson junctions of conventional superconductors, the Josephson current is carried by Cooper pairs and is $2\pi$ periodic in the phase difference applied to the junction. When the junction connects topological superconductors \cite{fu08,lutchyn10,oreg10,alicea11}, the coupling of Majorana bound states across the junction allows a Josephson current to flow by coherent transfer of single electrons, resulting in $4\pi$ periodicity in the phase difference. Robust $4\pi$ periodicity requires that time-reversal symmetry be broken through proximity coupling to a magnetic insulator or an applied magnetic field \cite{fu08}. A fractional Josephson effect can occur in time-reversal-symmetric junctions as a consequence of electron-electron interactions \cite{zhang2014,orth15}. In the limit of strong interactions, this $8\pi$-periodic effect can be understood in terms of domain walls carrying $\mathbb{Z}_4$ parafermions, enabling tunneling of $e/2$ quasiparticles between the superconductors. 

Recent experiments on superconductor -- quantum spin Hall -- superconductor junctions show intriguing evidence for $4\pi$-periodic Josephson currents. One experiment probes Shapiro steps and shows that the first Shapiro step is absent \cite{bocquillon16}.  A second experiment reports that the Josephson radiation emitted by a biased junction is also consistent with $4\pi$ periodicity \cite{deacon16}. These results are surprising as both experiments were performed without explicitly breaking time-reversal symmetry so that basic theory would predict a dissipative $2\pi$-periodic behavior when neglecting electron-electron interactions, or an $8\pi$-periodic behavior when taking interactions into account. 

These expectations are based on considering pristine quantum spin Hall Josephson junctions with a fully gapped bulk and a single helical channel propagating along its edges. Density modulations in actual quantum spin Hall samples are widely believed to induce puddles of electrons in addition to the helical edge channels \cite{varyrynen13}. When these puddles host an odd number of electrons, charging effects turn them into magnetic impurities which are exchange coupled to the helical edge channels. In this paper, we discuss the fractional Josephson effect in realistic quantum spin Hall Josephson junctions which include such magnetic impurities. 

The effects of magnetic impurities on quantum spin Hall edge channels have been intensively studied in the absence of superconductivity \cite{maciejko09,tanaka11,cheianov13,altshuler13}. In the high-temperature limit, a magnetic impurity induces backscattering between the Kramers pair of helical edge channels and
thus deviations from a quantized conductance in a two-terminal measurement. As the temperature is lowered, the impurity spin is increasingly Kondo screened by the helical edge channel and perfect conductance quantization is recovered when the temperature is low compared to the Kondo temperature $T_K$. In the presence of superconductivity, the Kondo effect is quenched by the superconducting gap $\Delta$ so that one may expect that magnetic impurities field more prominent consequences \cite{balatsky06}. Here, we assume that $T_K\ll\Delta$ so that we can safely neglect the effects of Kondo screening. 

We find that magnetic impurities alter the behavior of quantum spin Hall Josephson junctions qualitatively. The Josephson current becomes $8\pi$ periodic, replacing the dissipative $2\pi$-periodic effect in pristine junctions. This can be viewed as a variant of the $\mathbb{Z}_4$ Josephson effect. Indeed, unlike its classical counterpart, coupling to a quantum spin preserves time-reversal symmetry and interactions are effectively included through the local-moment formation. This is quite reminiscent of the ingredients of the $\mathbb{Z}_4$ fractional Josephson effect. Thus, our results show that this remarkable effect is considerably more generic than one might have previously thought. 

Moreover, the present setting emphasizes a remarkable mechanism for producing an $8\pi$-periodic fractional Josephson effect. As a result of the fermion parity anomaly \cite{fu09}, the spin of the helical edge effectively changes by $\hbar/2$ when the superconducting phase difference is advanced by $2\pi$. This adiabatically transmutes the combined spin of helical edge and magnetic impurity between half-integer and integer values, with their characteristically different behavior in the presence of time-reversal symmetry as described by the Kramers theorem.

{\em Quantum spin Hall Josephson junctions.---}We first review the Andreev spectrum of pristine quantum spin Hall Josephson junctions \cite{fu09}. Consider a quantum spin Hall edge with edge modes counterpropagating at velocity $v$, placed in between two superconductors at a distance $L$ whose phases differ by $\phi$. This junction is described by the Bogoliubov-de Gennes Hamiltonian 
\begin{equation}
    \mathcal{H}=vp\sigma_{z}\tau_{z}+\Delta(x) \tau_+ + \Delta^*(x)\tau_{-},
\end{equation}
where $\sigma_j$ and $\tau_j$ are Pauli matrices in spin and Nambu (particle-hole) space, respectively. 
The subgap spectrum as a function of $\phi$ is shown in Fig.\ 1. 

\begin{figure}[t]
    \centering
    \includegraphics[width=0.48\textwidth]{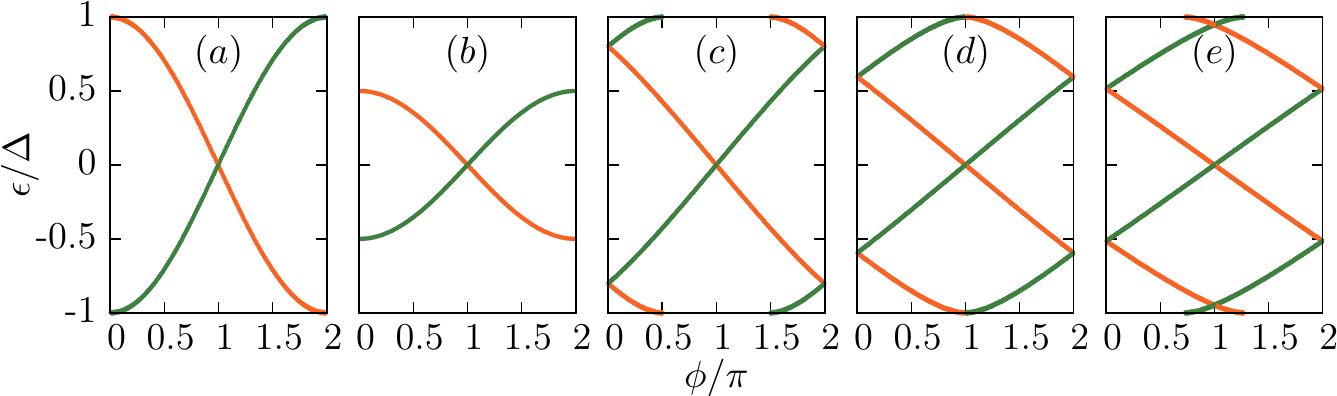}
        \caption{Andreev spectrum of quantum spin Hall Josephson junctions of different lengths.
          (a) $ L = 0$; (b) $ L = 0$ in the presence of backscattering due to a Zeeman field; 
        (c) $L = 0.8 \hbar v/\Delta$; (d) $ L=(\pi/2)\hbar v/\Delta$; (e) $ L = 2\hbar v/\Delta$. The green curves correspond to Andreev states consisting of a superposition of an up-spin electron and an Andreev-reflected hole. The orange curves are for the particle-hole conjugated states.}
    \label{fig:ABS}
\end{figure}

For short junctions ($L\to 0$), the subgap spectrum contains a particle-hole symmetric pair of Andreev states [see Fig.\ \ref{fig:ABS}(a)]. Both Andreev levels emanate from and merge into the quasiparticle continuum.   An applied bias voltage $V$ advances the phase difference at a rate $\dot\phi=2eV/\hbar$ and leads to the generation of continuum quasiparticles above the superconducting gap. These can diffuse away from the junction which causes dissipation. Thus, the junction exhibits an $ac$ Josephson effect with conventional frequency  and energy dissipation rate $(2\Delta) (\dot\phi/2\pi)$.

The dissipative nature of the Josephson effect is closely related to the absence of backscattering. When introducing backscattering into the junction by breaking time-reversal symmetry through an applied magnetic field or proximity coupling to a magnetic insulator, the Andreev levels no longer merge with the quasiparticle continuum [see Fig.\ \ref{fig:ABS}(b)]. Now, the quasiparticles generated by the advancing phase difference remain at subgap energies and localized at the junction, which quenches dissipation in the small-voltage limit \cite{fu08}. Moreover, the $ac$ Josephson effect occurs at half the conventional frequency, i.e., at $eV/\hbar$, as fermion number parity is conserved. Indeed, the level crossing at $\phi=\pi$ is protected by fermion number parity so that the individual Andreev levels are $4\pi$ periodic in the phase difference $\phi$. This can be viewed as a consequence of the fermion parity anomaly (see \cite{suppl} for more details): As a result of the quantum spin Hall effect, the parity of the fermion number of the edge changes when the superconducting phase difference is advanced by $2\pi$, requiring a phase change of $4\pi$ for a full period. 

Additional subgap levels appear for longer junctions, see Figs.\ \ref{fig:ABS}(c) and (d). The level crossings in these spectra are not only controlled by fermion number parity, but also by time-reversal symmetry. While time reversal is broken by the phase difference across the junction (causing a nonzero Josephson current to flow), it remains unbroken when $\phi$ is an integer multiple of $\pi$. 

 {\em Coupling to magnetic impurity.---}We now consider the coupling of the edge channel to a magnetic impurity with spin S. Generically, disorder in conjunction with the strong spin-orbit coupling will remove any symmetry other than time reversal which we assume to be broken only by the applied superconducting phase difference. Thus, we focus on the general Hamiltonian 
\begin{equation}
   H_S = \sum_{\alpha,\beta} J_{\alpha\beta} \hat S^\alpha \hat \sigma^\beta(0) + \sum_\alpha D_\alpha (\hat S^\alpha)^2
   \label{H}
\end{equation}
for the impurity spin $\hat{\mathbf S}$. The first term describes the exchange coupling between the impurity spin and the helical edge, with $\hat\sigma^\alpha(0) = \sum_{i,j} \psi_i^\dagger(0) (\sigma^\alpha)_{ij}\psi_j(0)$ denoting the local spin density of the edge at the position $x=0$ of the impurity. The operator $\psi_i(x)$ annihilates an electron with spin projection $i$ at position $x$. The second term describes a single-ion anisotropy of the impurity spin with strengths $D_\alpha$. Time reversal implies that the exchange couplings are real, but otherwise arbitrary. 

\begin{figure}[t]
    \centering
    \includegraphics[width=0.45\textwidth]{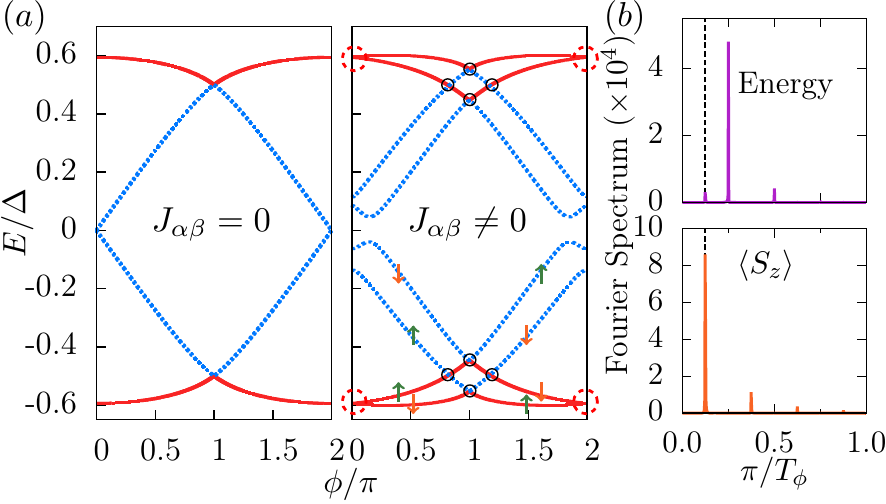}
    \caption{(a) Generic many-body spectrum for the quantum spin Hall Josephson junction ($L=(\pi/2)\hbar v/\Delta$) without (left) and with coupling to the impurity spin (right; for parameters see Ref.~\cite{suppl}). The red solid and blue dashed curves indicate even and odd fermion number parity, respectively. The discontinuity in fermion number parity at $\phi=\pi$ originates from the merging of Andreev levels with the continuum, see Fig.\ \ref{fig:ABS}(d). The crossings at and near  $\phi = \pi$ (black circles) are between states of opposite fermion number parity. The crossings between states with even fermion number parity at $\phi=0$ and $2\pi$ (red dashed circles) are protected by time reversal. The arrows indicate the impurity-spin polarization along the $z$-axis. (b) Fourier transforms of the many-body ground state energy (equivalently: Josephson current) (upper panel) and of the expectation value of the impurity spin $\braket{S_{z}}$ (lower panel) as a function of the phase difference $\phi$. The $8\pi$-periodic harmonics are indicated by the vertical dashed lines.}
    \label{fig:S=half}
\end{figure}

{\em Josephson effect.}---Analyzing the Josephson effect of the quantum spin Hall edge channel coupled to the magnetic impurity is greatly simplified by the discrete nature of the subgap spectrum. For definiteness, consider an intermediate-length junction whose subgap spectrum has exactly two positive-energy subgap states $\epsilon_n(\phi)$ ($n=1,2$)  at all values of the phase difference as in Fig.\ \ref{fig:ABS}(d). (This convenient choice is used in our numerical illustrations but not essential for our results.) Then, we can analyze the low-energy (many-body) spectrum of the junction in the finite-dimensional space of low-energy states spanned by the product of occupation states of the two subgap Bogoliubov quasiparticles (yielding four basis states) and the $2S+1$ spin states of the spin-$S$ impurity. The low-energy many-body spectrum effectively decouples from the quasiparticle continuum when the Kondo temperature is small compared to the superconducting gap \cite{foot}. The corresponding Hamiltonian is readily derived by retaining only the contributions of the two positive-energy subgap Bogoliubov operators $\gamma_n$ to the edge-state electron operators (see \cite{suppl} for details). In this limit, the total Hamiltonian can be approximated as $H=H_e+H_S$ with 
\begin{equation}
   H_e = \sum_n \epsilon_{n}(\phi) \left(\gamma_{n}^\dagger \gamma_n - \frac{1}{2} \right)
   \label{eq:Hbare}
\end{equation}
the Hamiltonian of the bare edge. 

Consider coupling the quantum spin Hall edge states to a spin-1/2 impurity. Figure \ref{fig:S=half}(a) shows the many-body spectrum of $H_e$ in Eq.\ (\ref{eq:Hbare}), i.e., of the bare edge (left panel), and of $H = H_e+H_S$ for a generic choice of exchange couplings $J_{\alpha\beta}$ (right panel). The spectrum of the coupled edge is best understood by analyzing the nature of the degeneracies at phase differences equal to integer multiples of $\pi$. The degeneracies at and near $\phi=\pi$ are protected by fermion number parity. Here, level crossings occur between states with even and odd occupations of the Bogoliubov quasiparticles of the edge. In contrast, the level crossings at $\phi=0$ and $\phi=2\pi$ occur between states of the same fermion number parity and are Kramers degeneracies reflecting time-reversal symmetry. 

In the present system, a Kramers degeneracy appears when the Bogoliubov quasiparticles $\gamma_n$ of the edge are either both empty or both occupied, leading to a half-integer spin of the combined system of edge and impurity. Specifically, the lower (higher) energy crossing in Fig.\ \ref{fig:S=half}(a) corresponds to states in which the quasiparticle states are both empty (occupied). Away from $\phi=0$ and $2\pi$, time reversal is broken and the Kramers degeneracies are lifted. This interpretation is corroborated by further
restricting the Hamiltonian $H$ for small $\phi$ to the low-energy subspace of empty quasiparticle states. In this limit, the spin density $\hat\sigma^\alpha(0)$ of the edge only has a nonzero $z$ component $\hat\sigma^z(0)=-\epsilon\phi/[2\hbar v(1+\kappa L)^2]$ and the Hamiltonian simplifies to 
\begin{equation}
H \simeq -\sum_\alpha B^{\alpha}S^\alpha + {\rm const}
\end{equation}
with the effective Zeeman field ${\bf B} = [\epsilon\phi/2\hbar v(1+\kappa L)^2] \sum_\alpha J_{\alpha z} {\bf \hat e}_\alpha$. Here, we use the subgap energy $\epsilon=\Delta\cos(\epsilon L/(\hbar v))$ and $\kappa = \sqrt{\Delta^2-\epsilon^2}/(\hbar v)$.   

The four nondegenerate states at intermediate energies for $\phi=0$ [see Fig.\ \ref{fig:S=half}(a)] have overall single occupation of the quasiparticle states, leading to a combined edge-impurity system with integer spin. Unlike in the odd-integer spin case, time reversal does not enforce a degeneracy of the many-body spectrum in this case. Writing the Hamiltonian for small $\phi$ in this subspace using the basis $|\uparrow\rangle = \gamma_1^\dagger |{\rm gs}\rangle$ and $|\downarrow\rangle = \gamma_2^\dagger |{\rm
gs}\rangle$ (with the junction ground state $\ket{\rm gs}$ such that $\gamma_1 |{\rm gs}\rangle=\gamma_2 |{\rm gs}\rangle=0$) for the states of the edge (with corresponding Pauli matrices $\rho_\alpha$), we find the effective Hamiltonian
\begin{equation}
   H\simeq \frac{\kappa}{2(1+\kappa L)} \left[\sum_\alpha J_{\alpha +}S^\alpha \rho_+ + {\rm h.c.}\right]
\end{equation}
Generically, this Hamiltonian has no degeneracies. 

With this understanding, the many-body spectrum in Fig.\ \ref{fig:S=half}(a) reveals a remarkable fact: Adiabatically advancing the superconducting phase difference by $2\pi$ connects the low-energy Kramers doublet at $\phi=0$ to states of the totally lifted spin quartet at $\phi=2\pi$. Thus, adiabatic quantum dynamics changes the total spin of the edge-impurity system between half-integer and integer values. This spin transmutation is a direct consequence of the fermion parity anomaly (see also \cite{suppl}): As the phase difference changes by $2\pi$, the fermion number parity of the edge changes by virtue of the quantum spin Hall effect. Consequently, also the spin of the edge changes by $\hbar/2$. This change in spin has important consequences for the periodicity of the Josephson effect. Indeed, adiabatically following the energy levels in Fig.\ \ref{fig:S=half}(a), we find that they are $8\pi$ periodic, corresponding to an $ac$ Josephson frequency of $eV/2\hbar$. Due to the spin transmutation, the system passes through successive Kramers degeneracies only after advancing the superconducting phase difference by $4\pi$, requiring a phase change of $8\pi$ for completing a full period.  Note that starting with the ground state at $\phi=0$, the many-body state remains well below the quasiparticle continuum for all $\phi$, so that the $ac$ Josephson effect is nondissipative at a sufficiently small bias. 

The polarization of the impurity spin varies with the superconducting phase difference in an $8\pi$-periodic manner. When adiabatically varying $\phi$, the spin orientation remains unchanged at the Kramers crossings and flips in the vicinity of the avoided crossings where the edge-impurity system is in an integer-spin state.  This variation of the spin with $\phi$ is illustrated in Fig.\ \ref{fig:S=half}(a).

\begin{figure}[t]
    \centering
    \includegraphics[width=0.45\textwidth]{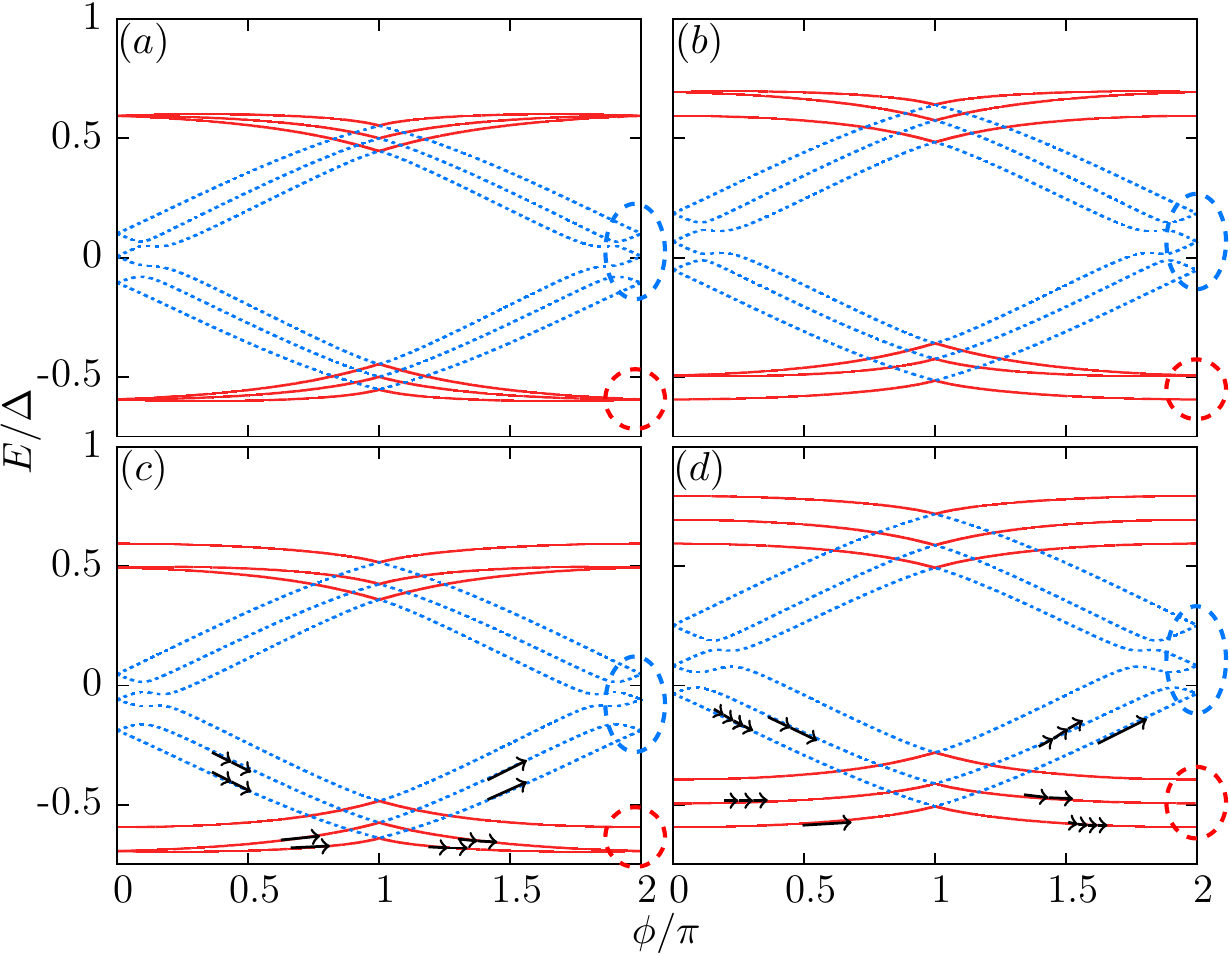}
    \caption{Many-body spectrum for a quantum spin Hall edge coupled to an $S=1$ impurity (for explicit parameters, see \cite{suppl}). The red solid and blue dashed curves correspond to many-body states with even and odd fermion number parity, respectively.  Spectra correspond to (a) vanishing single-ion anisotropy, (b) easy-plane anisotropy $D_z>0$, (c) easy-axis anisotropy $D_{z}<0$, and (d) generic single-ion anisotropy with $D_{x},D_{y},D_{z}\neq 0$. The degeneracies at $\phi=2\pi$ (blue dashed circles)
and their partners at $\phi=0$ are Kramers degeneracies. Red circles highlight degeneracies which are lifted by generic single-ion anisotropy. The number of arrows indicates subsequent $2\pi$ periods when adiabatically advancing $\phi$.}
    \label{fig:S=1}
\end{figure}

These results for $S=1/2$ impurities persist for higher-spin impurities.  Results for an $S=1$ impurity are shown in Fig.\ \ref{fig:S=1}. Panel (d) shows results for generic values of $J_{\alpha\beta}$ and $D_\alpha$. Unlike in the $S=1/2$ case, the low-energy states now have integer spin and are nondegenerate, while the intermediate-energy states have half-integer spin and are Kramers degenerate at $\phi=0$ and $2\pi$. Nevertheless, the $8\pi$ periodicity remains intact. 

Different periodicities occur for nongeneric $D_\alpha$. Without single-ion anisotropy [see Fig.\ \ref{fig:S=1}(a)], the spectrum does not decouple from the quasiparticle continuum and the Josephson effect becomes dissipative and $2\pi$ periodic. The same results for easy-plane anisotropy, with one of the single-ion
anisotropies being positive and the others equal to zero, see Fig.\ \ref{fig:S=1}(b). Finally, easy-axis anisotropy makes the junction nondissipative and  $4\pi$ periodic as shown
in Fig.\ \ref{fig:S=1}(c). 

{\em Discussion.---}We find that generically, coupling to a magnetic impurity makes the Josephson effect in quantum spin Hall systems $8\pi$ periodic, corresponding to a frequency $eV/2\hbar$ of the $ac$ Josephson effect. The $8\pi$ periodicity  relies only on time-reversal symmetry, the parity anomaly, and the absence of fine tuning such as the absence of interactions or the presence of spin conservation. It can be thought of as resulting from the coupling of $\mathbb{Z}_4$ parafermions across the junction.

This general conclusion requires two comments. First, the $8\pi$-periodic Josephson current may not be the dominant Fourier component in experiment. Indeed, as is evident from Fig.\ \ref{fig:S=half}, the $8\pi$-periodic cycle consists of two rather similar $4\pi$ sections. The splitting between the two sections is controlled by the exchange coupling. When the exchange splitting is small compared to the superconducting gap, the dominant Fourier component of the Josephson current is $4\pi$ periodic. This is shown in Fig.\ \ref{fig:S=half}(b), together with the Fourier components of the impurity spin polarization which has a dominant $8\pi$-periodic harmonic. It is interesting to note that this result for the Josephson current is different from the realization of the $\mathbb{Z}_4$ Josephson effect discussed by Zhang and Kane \cite{zhang2014} which has a dominant $8\pi$-periodic Fourier component. 

Second, our results so far consider only the electronic system. Coupling to other degrees of freedom such as phonons or the electromagnetic environment introduces inelastic relaxation processes which may crucially affect the experimentally observed periodicity. While relaxation between states of opposite fermion number parity may be slow, parity-conserving relaxation processes should be considerably more efficient. Observation of the $8\pi$ periodicity requires that the latter relaxation processes be slow compared to the time in which the $8\pi$ cycle is traversed. Indeed, the two $4\pi$ sections of the $8\pi$ cycle involve states of the same fermion number parity. Thus, the system always remains in the lower-energy state if the cycle is traversed slowly on the time scale of parity-conserving relaxation processes. This makes the observed Josephson effect $4\pi$ rather than $8\pi$ periodic. 

\begin{table}[t]
    \begin{center}
    \begin{tabular}{|C{1.5cm}|C{2cm}|C{2.4cm}|C{2.4cm}|}
        \hline 
        Josephson effect & $S=0$, $B=0$ & any $S$, $B\neq0$ & $S\neq0$, $B=0$\tabularnewline
        \hline 
        $dc$  & $2\pi$ & $2\pi$ & $2\pi$\tabularnewline
        \hline 
        $ac$  & diss., $2\pi$ & non-diss., $4\pi$  & non-diss., $8\pi$\tabularnewline
        \hline 
    \end{tabular}
\end{center}
 \caption{\label{tab:Josephson} Generic Josephson periodicities of a quantum-spin-Hall-based junction coupled to a spin-$S$ impurity/quantum dot, with or without Zeeman field $B$. The $dc$ periodicities assume full equilibration including fermion number parity. The $ac$ results assume that fermion number parity is conserved on the relevant time scale. Note that the $8\pi$-periodic $ac$ current may have a large $4\pi$-periodic component, while the periodicity of the dot spin is robustly $8\pi$, see Figs.~\ref{fig:S=half}(b) and (c).}
 \label{table}
\end{table}

It is interesting to compare these results to the recent experiments on quantum spin Hall junctions which observe Shapiro steps and Josephson radiation consistent with $4\pi$ periodicity \cite{bocquillon16,deacon16}. Our results provide an intriguing scenario that is consistent with these observations. However, this is not the only explanation of a $4\pi$-periodic Josephson effect in this system. An alternative scenario considers relaxation processes in a pristine quantum spin Hall junction. Consider an intermediate-length junction with at least two positive-energy Andreev states for any phase difference. When both of these Andreev states are occupied, the two quasiparticles can relax inelastically by recombining into a Cooper pair. Two positive-energy quasiparticles are created every time the phase difference advances by $4\pi$. Thus, if recombination into a Cooper pair is an efficient process, one would also observe a $4\pi$-periodic Josephson effect. It is an interesting problem to devise experiments which distinguish between these alternative scenarios. Such efforts may benefit from the considerable recent progress in directly probing the subgap spectrum of Josephson junctions by microwave spectroscopy and switching current measurements \cite{bretheau13,bretheau13nature,yale15,peng2016,delft16}. 

Finally, our results suggest probing the Josephson effect of a quantum spin Hall edge which is intentionally coupled to a quantum dot. Such a setup would allow one to tune the quantum dot in and out of the local moment regime and to control the exchange coupling between dot and edge. In addition to the Josephson periodicity, such a setup might provide access to the $8\pi$ periodicity of the impurity spin (see Table \ref{table})  and would be a promising setup for detecting $\mathbb{Z}_4$ parafermions.

{\em Acknowledgments.---}We acknowledge financial support by the Deutsche Forschungsgemeinschaft (CRC 183 as well as Priority Program 1666), the Minerva Foundation, the Alexander-von-Humboldt Foundation, and NSF Grant DMR-1603243. Part of this work was performed at the Aspen Center for Physics, which is supported by National Science Foundation grant PHY-1066293.

\newpage
\setcounter{equation}{0}
\begin{widetext}
\section*{Supplemental Material}

\section*{Fermion number parity}

For the benefit of general readers, we include a brief review of the concept of fermion number parity and its application to topological Josephson junctions. Pairing Hamiltonians (i.e., Hamiltonians describing superconductors within mean-field theory) include terms which change the number of particles of the system by two. This can be thought of as describing the addition or removal of Cooper pairs from the system. As a result, these Hamiltonians break particle number conservation but preserve particle number modulo two. Consequently, systems with even and odd numbers of electrons decouple and fermion number parity  is a good quantum number. One refers to systems with even (odd) electron number as having even (odd) fermion number parity.

The Hamiltonian of a Josephson junction based on a quantum spin Hall edge also conserves fermion number parity. We can label the many body eigenstates of the system by fermion number parity and states with different fermion number parity cannot anticross, but must exhibit a true level crossing. This explains the level crossings in the many-body spectra near a phase difference of $\pi$, see Fig.\ 2(a) of the main text. 

While the spectrum behaves as if fermion number parity is a good quantum number, it is not conserved in the quantum dynamics. This is referred to as an anomaly (or more specifically the fermion parity anomaly): While the fermion number parity is conserved classically, it is not in the quantum dynamics. In the absence of an anomaly, the fermion number parity has to remain unchanged when changing any parameter in the (fermion-parity-conserving) Hamiltonian. As a corollary, this also implies that the spin has to remain integer or half-integer at all times, and transmutation between integer and half-integer spin is forbidden.

Nevertheless, in the present case, the system transmutes between integer and half-integer spin (and even and odd fermion parity) when the superconducting phase difference is advanced by $2\pi$. When the phase difference advances adiabatically, the system follows a specific eigenstate. Due to the crossings near $\phi=\pi$ protected by fermion parity, the system passes, say, from a half-integer Kramers doublet at zero phase difference to a state which is part of a completely lifted integer-spin quartet at $\phi=2\pi$, see Fig.\ 2(b) of the main text. Thus, the adiabatic quantum dynamics does indeed violate fermion parity conservation.

Physically, the fermion parity anomaly can be understood as follows. In a Corbino geometry, changes in the superconducting phase difference can be effected by changing the magnetic flux piercing the hole of the Corbino disk. Changes in flux induce an azimuthal electric field circulating around the Corbino disk. By virtue of the quantum spin Hall effect, this azimuthal electric field drives a radial spin Hall current. It is a simple exercise to compute the total spin change of the inner and outer edges of the Corbino disk when the superconducting phase is advanced by $2\pi$ and one finds that it is exactly $\hbar/2$. It is by this mechanism that the spin of the edge states transmutes between integer and half integer spins. 

\section*{Andreev Bound States}
The Hamiltonian for the quantum spin Hall Josephson junction takes the form
$H=\frac{1}{2}\Psi^\dagger \mathcal{H} \Psi$ with Nambu spinor $\Psi=(\psi_{\uparrow},\psi_{\downarrow},\psi_{\downarrow}^{\dagger},-\psi_{\uparrow}^{\dagger})^{T}$ in terms of electron operators
and the Bogoliubov-de Gennes Hamiltonian 
\begin{equation}
    \mathcal{H}=vp\sigma_{z}\tau_{z}+\Delta(x)\tau_{x},
\end{equation}
where $x$ ($p$) denotes the coordinate (momentum) along the quantum spin Hall insulator edge, $v$ is the edge-mode velocity, and $\sigma_j$ and $\tau_j$ are Pauli matrices in spin and Nambu (particle-hole) space, respectively. The proximity-induced superconducting gap
\begin{align}
    \Delta(x)=&\Delta\left[ \theta(-x-L/2)+e^{i\phi\tau_z}\theta(x-L/2)\right] \nonumber \\
    &= \Delta\theta(\abs{x}-\frac{L}{2}) e^{i\varphi(x)\tau_z} 
\end{align}
has strength $\Delta>0$ and a phase difference $\phi$ across the junction region of length $L$.
We have introduced a spatially dependent phase  
\begin{equation}
    \varphi(x)= \frac{\phi}{L}(x+\frac{L}{2})\theta(\frac{L}{2}-\abs{x}) + \theta(x-\frac{L}{2})\phi.
\end{equation}
for convenience. We set $\hbar=1$ in this supplemental material.

We introduce a local gauge transformation $U=e^{i\varphi(x)\tau_z/2}$ to eliminate the spatial dependence of the superconducting phase, 
and obtain the transformed Hamiltonian
\begin{equation}
    U^{\dagger}\mathcal{H}U= -iv\partial_{x}\sigma_{z}\tau_{z} + \frac{v\varphi^\prime(x)\sigma_z}{2} + \Delta\theta(\abs{x}-\frac{L}{2})\tau_x,
\end{equation}
where the prime denotes a derivative with respect to $x$. We will denote $U^{\dagger}\mathcal{H}U$ as $\mathcal{H}$ in the following.

To solve for the Andreev bound states, we follow the approach detailed in Ref. \cite{peng2016} to rearrange the Bogoliubov-de Gennes equation ${\cal H}\psi = \epsilon\psi$ as
\begin{equation}
    i\frac{\partial\psi}{\partial x} = -\frac{1}{v}\sigma_z\tau_z\left[ \epsilon-\Delta\theta(\abs{x}-\frac{L}{2})\tau_x -\frac{v\varphi^\prime(x)}{2}\sigma_z    \right] \psi.
\end{equation}
The solution can be written as $\psi(x) = U(x,x_0) \psi(x_0)$ in terms of the state at some reference point $x_0$. In particular, 
we have 
\begin{equation}
    U(\frac{L}{2},-\frac{L}{2}) = \exp\left\{\frac{i}{v} \sigma_z \tau_z \int_{-\frac{L}{2}}^{\frac{L}{2}}dx'\, 
    \left[\epsilon - \frac{v\varphi^\prime(x')}{2}\sigma_z \right] = \exp\left(\frac{iEL}{v}\sigma_z\tau_z-\frac{\phi}{2}\tau_z\right)
\right\} 
\end{equation}
which connects the states $\psi(L/2) = U(L/2,-L/2) \psi(-L/2)$. We match the properly decaying solutions of the Bogoliubov de-Gennes equation on the left and right of the junction, and obtain the bound state wave functions
in the two spin sectors:
\begin{equation}
    \Psi_{\uparrow}(x)=\left(\begin{array}{c}
        a_{\uparrow}A_{\uparrow}\\
        0\\
        A_{\uparrow}\\
        0
    \end{array}\right)e^{\kappa_{\uparrow}(x+\frac{L}{2})}\theta(-x-\frac{L}{2})+\left(\begin{array}{c}
        a_{\uparrow}A_{\uparrow}e^{-i(\frac{\phi}{2}-\frac{\epsilon}{v}L)}\\
        0\\
        A_{\uparrow}e^{i(\frac{\phi}{2}-\frac{\epsilon}{v}L)}\\
        0
    \end{array}\right)e^{-\kappa_\uparrow(x-\frac{L}{2})}\theta(x-\frac{L}{2})+\left(\begin{array}{c}
        a_{\uparrow}A_{\uparrow}e^{-i(\frac{\phi}{2L}-\frac{\epsilon}{v})(x+\frac{L}{2})}\\
        0\\
        A_{\uparrow}e^{i(\frac{\phi}{2L}-\frac{\epsilon}{v})(x+\frac{L}{2})}\\
        0
    \end{array}\right)\theta(\frac{L}{2}-\abs{x})
\end{equation}
and
\begin{equation}
    \Psi_{\downarrow}(x)=\left(\begin{array}{c}
        0\\
        A_{\downarrow}\\
        0\\
        a_{\downarrow}A_{\downarrow}
    \end{array}\right)e^{\kappa_\downarrow(x+\frac{L}{2})}\theta(-x-\frac{L}{2})+\left(\begin{array}{c}
        0\\
        A_{\downarrow}e^{-i(\frac{\phi}{2}+\frac{\epsilon}{v}L)}\\
        0\\
        a_{\downarrow}A_{\downarrow}e^{i(\frac{\phi}{2}+\frac{\epsilon}{v}L)}
    \end{array}\right)e^{-\kappa_\downarrow(x-\frac{L}{2})}\theta(x-\frac{L}{2})+\left(\begin{array}{c}
        0\\
        A_{\downarrow}e^{-i(\frac{\phi}{2L}+\frac{\epsilon}{v})(x+\frac{L}{2})}\\
        0\\
        a_{\downarrow}A_{\downarrow}e^{i(\frac{\phi}{2L}+\frac{\epsilon}{v})(x+\frac{L}{2})}
    \end{array}\right)\theta(\frac{L}{2}-\abs{x})
\end{equation}
where
\begin{equation}
  \label{eq:A_updown}
    a_{\sigma} = \frac{\epsilon}{\Delta} - i\frac{\sqrt{\Delta^{2} - \epsilon_{\sigma}^2}}{\Delta},\qquad \abs{A_{\sigma}}^{2}=\frac{\kappa_{\sigma}}{2(1+L\kappa_{\sigma})},\qquad\kappa_{\sigma}=\frac{\sqrt{\Delta^{2}-\epsilon_{\sigma}^{2}}}{v}
\end{equation}
with $\sigma = \uparrow, \downarrow$, and $\epsilon_{\sigma}$ is the positive eigenvalue in each spin sector given by the relation
\begin{equation}
    \epsilon_{\uparrow}/\Delta=\begin{cases}
        \cos(\frac{\phi}{2}-\frac{\epsilon_\uparrow}{v}L) & \sin(\frac{\phi}{2}-\frac{\epsilon_\uparrow}{v}L)<0\\
        -\cos(\frac{\phi}{2}-\frac{\epsilon_\uparrow}{v}L) & \sin(\frac{\phi}{2}-\frac{\epsilon_\uparrow}{v}L)>0
    \end{cases}
    \label{eq:up_spin}
\end{equation}
and
\begin{equation}
    \epsilon_{\downarrow}/\Delta=\begin{cases}
        \cos(\frac{\phi}{2}+\frac{\epsilon_\downarrow}{v}L) & \sin(\frac{\phi}{2}+\frac{\epsilon_\downarrow}{v}L)>0\\
        -\cos(\frac{\phi}{2}+\frac{\epsilon_\downarrow}{v}L) & \sin(\frac{\phi}{2}+\frac{\epsilon_\downarrow}{v}L)<0
    \end{cases}.
    \label{eq:down_spin}
\end{equation}

For $\phi$ around $2n\pi,n\in\mathbb{Z}$, solutions in both spin sectors  can exist simultaneously for $L>0$. We will consider the situation when at most one solution in each spin sector exists, and write the subgap effective Hamiltonian as
\begin{equation}
    H_{e}=\sum_{\sigma}\epsilon_{\sigma}(\phi)(\gamma_{\sigma}^{\dagger}\gamma_{\sigma}-\frac{1}{2}),
\end{equation}
where $\gamma_{\sigma}$ is the Bogoliubov quasiparticle annihilation operator for the Andreev bound state with spin $\sigma$. 
When projected onto the subspace spanned by the subgap Andreev bound states, the electron annihilation operators
for both spins can be written approximately as
\begin{gather}
\psi_{\uparrow}=a_{\uparrow}A_{\uparrow}e^{-ik_{\uparrow}L/2}\gamma_{\uparrow}-a_{\downarrow}^{*}A_{\downarrow}^{*}e^{-ik_{\downarrow}L/2}\gamma_{\downarrow}^{\dagger}  \nonumber \\
\psi_{\downarrow}=A_{\downarrow}e^{-ik_{\downarrow}L/2}\gamma_{\downarrow}+A_{\uparrow}^{*}e^{-ik_{\uparrow}L/2}\gamma_{\uparrow}^{\dagger}\nonumber  \\
k_{\uparrow,\downarrow}=\frac{\varphi}{2L}\mp\frac{\epsilon_{\sigma}}{v}.
\end{gather}

\section*{Coupling of an edge channel to a magnetic imurity}
\begin{figure}[h]
  \includegraphics[width=0.5\textwidth]{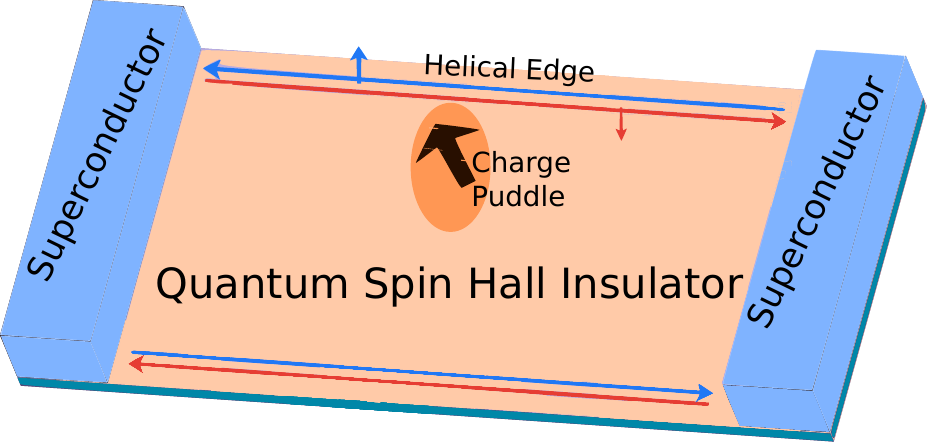}
  \caption{\label{fig.QSHJ} Setup for the quantum spin Hall Josephson junction in which the helical edge state is coupled to a charge puddle formed by a potential variation in the bulk. Due to charging effects, the charge puddle may effectively act as a magnetic impurity as indicated by the black arrow in the figure. The left/right moving electrons of the helical edges with opposite spins are depicted in different colors. }
\end{figure}

Consider the Hamiltonian describing the coupling of the edge channel to a magnetic impurity with spin S (see setup in Fig.~\ref{fig.QSHJ})
\begin{equation}
       H_S = \sum_{\alpha,\beta} J_{\alpha\beta} \hat S^\alpha \hat \sigma^\beta(0) + \sum_\alpha D_\alpha (\hat S^\alpha)^2,
\end{equation}
in which the spin density of the helical edge can be written in term of the Bogoliubov operators
\begin{gather}
    \hat{\sigma}^{+} = \psi_{\uparrow}^{\dagger}\psi_{\downarrow} = \left(a_{\uparrow}^{*}e^{i\Delta kL/2}+a_{\downarrow}e^{-i\Delta kL/2}\right)A_{\uparrow}^{*}A_{\downarrow}\gamma_{\uparrow}^{\dagger}\gamma_{\downarrow} \nonumber \\
    \hat{\sigma}^{-} = \psi_{\downarrow}^{\dagger}\psi_{\uparrow}=\left(a_{\uparrow}e^{-i\Delta kL/2}+a_{\downarrow}^{*}e^{i\Delta
    kL/2}\right)A_{\uparrow}A_{\downarrow}^{*}\gamma_{\downarrow}^{\dagger}\gamma_{\uparrow} \nonumber \\
    \hat{\sigma}^{x} = \left(\hat{\sigma}^{+} + \hat{\sigma}^{-}\right)/2, \quad \hat{\sigma}^{y} = \left(\hat{\sigma}^{+} + \hat{\sigma}^{-}\right)/2i \nonumber \\
    \hat{\sigma}^{z}=\psi_{\uparrow}^{\dagger}\psi_{\uparrow}-\psi_{\downarrow}^{\dagger}\psi_{\downarrow}=\left(\abs{A_{\uparrow}}^{2}(2\gamma_{\uparrow}^{\dagger}\gamma_{\uparrow}-1)-\abs{A_{\downarrow}}^{2}(2\gamma_{\downarrow}^{\dagger}\gamma_{\downarrow}-1)\right)+(a_{\uparrow}^{*}a_{\downarrow}^{*}e^{i\Delta
    kL/2}-e^{-i\Delta kL/2})A_{\uparrow}^{*}A_{\downarrow}^{*}\gamma_{\downarrow}^{\dagger}\gamma_{\uparrow}^{\dagger} \nonumber \\
    +(a_{\uparrow}a_{\downarrow}e^{-i\Delta kL/2}-e^{i\Delta
    kL/2})A_{\uparrow}A_{\downarrow}\gamma_{\uparrow}\gamma_{\downarrow},
\end{gather}
where 
\begin{equation}
    \Delta k=k_{\uparrow} - k_{\downarrow}=-\frac{\epsilon_{\uparrow}+\epsilon_{\downarrow}}{v}.
\end{equation}

Note that we can also write
\begin{gather}
    a_{\uparrow}=\Sgn{\sin(\frac{\epsilon_{\uparrow}L}{v}-\frac{\phi}{2})}e^{-i\frac{\epsilon_{\uparrow}L}{v}}e^{i\phi/2},\quad
    a_{\downarrow}=\Sgn{\sin(\frac{\epsilon_{\downarrow}L}{v}+\frac{\phi}{2})}e^{-i\frac{\epsilon_{\downarrow}L}{v}}e^{-i\phi/2},
\end{gather}
then we have 

\begin{gather}
    \hat{\sigma}^{+}=\left(\Sgn{\sin(\frac{\epsilon_\uparrow L}{v}-\frac{\phi}{2})}+\Sgn{\sin(\frac{\epsilon_\downarrow L}{v}+\frac{\phi}{2})}\right)e^{i\frac{(\epsilon_\uparrow -
    \epsilon_\downarrow) L}{2v}}e^{-i\phi/2}A_{\uparrow}^{*}A_{\downarrow}\gamma_{\uparrow}^{\dagger}\gamma_{\downarrow}\\
    \hat{\sigma}^{-}=\left(\Sgn{\sin(\frac{\epsilon_\uparrow L}{v}-\frac{\phi}{2})}+\Sgn{\sin(\frac{\epsilon_\downarrow L}{v}+\frac{\phi}{2})}\right)
    e^{-i\frac{(\epsilon_\uparrow-\epsilon_\downarrow)L}{2v}}e^{i\phi/2}A_{\uparrow}A_{\downarrow}^{*}\gamma_{\downarrow}^{\dagger}\gamma_{\uparrow}   
\end{gather}
\begin{align}
    \hat{\sigma}^{z}&=\left(\abs{A_{\uparrow}}^{2}(2\gamma_{\uparrow}^{\dagger}\gamma_{\uparrow}-1)-\abs{A_{\downarrow}}^{2}(2\gamma_{\downarrow}^{\dagger}\gamma_{\downarrow}-1)\right)
    \nonumber \\
    &+\left[\Sgn{\sin(\frac{\epsilon_\uparrow L}{v}-\frac{\phi}{2})}\Sgn{\sin(\frac{\epsilon_\downarrow L}{v}+\frac{\phi}{2})}-1\right]\left(e^{i\frac{(\epsilon_\uparrow +\epsilon_\downarrow)L}{2v}}
    A_{\uparrow}^{*}A_{\downarrow}^{*}\gamma_{\downarrow}^{\dagger}\gamma_{\uparrow}^{\dagger}+e^{-i\frac{(\epsilon_\uparrow+\epsilon_\downarrow)L}{2v}}A_{\uparrow}A_{\downarrow}
    \gamma_{\downarrow}^{\dagger}\gamma_{\uparrow}^{\dagger}  \right). 
\end{align}

\begin{figure}[h]
  \includegraphics[width=0.5\textwidth]{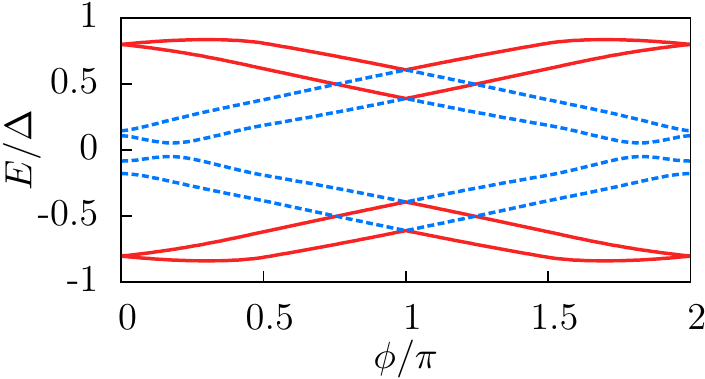}
  \caption{\label{fig.L_finite}Generic many body spectrum for the quantum spin Hall Josephson junction
coupled to a spin-$1/2$ impurity with $\Delta L/v=0.8$.  The red solid and blue dashed curves indicate that the
corresponding many-body states have even and odd fermion parity, respectively. The coupling between the impurity and the edge and the single-ion anisotropy are chosen to be the same as in Fig.\ 2 of the main text, see Eq.\ (\ref{eq:J}).}
\end{figure}

\section*{Parameters for the coupling matrix used in Figs.\ 2 and 3 of the main text}

In Fig.\ 2 of the main text, the coupling matrix between edge and spin-1/2 impurity was chosen as 
\begin{equation}
    \frac{J^{S=1/2}}{2\pi v}=
    \left(
    \begin{array}{ccc}
       0.2041 & 0.124268 & 0.33448 \\
        0.197511 & -0.185256 & 0.0733386 \\
         0.394004 & 0.0849569 & 0.288134 \\
       \end{array}
       \right)
      \label{eq:J}
\end{equation}
The single-ion anisotropy was set to zero since it is only a constant shift in energy for spin-$1/2$ impurities.
A similar figure for a shorter junction with the same parameters is shown in Fig.~\ref{fig.L_finite}.

In Fig.~3 of the main text, the coupling matrix between edge and spin-1 impurity was chosen as $J^{S=1} = J^{S=1/2}/2$. In Fig.~3(b), $D_z = 0.1\Delta$. In Fig.~3(c), $D_z=-0.1\Delta$. In Fig.~3(d), $D_x=0.3, D_y=0.2, D_z=0.1$.

\section*{Analysis around $\phi=0$}
In the case $\Delta L/v \in [0,\pi/2]$, and for $\phi$ close to $0$ we have
\begin{equation}
  \Sgn \sin(\frac{\epsilon_{\downarrow}L}{v} - \frac{\phi}{2} ) = \Sgn \sin(\frac{\epsilon_{\uparrow}L}{v} + \frac{\phi}{2})=1.
  \label{eq:condition}
\end{equation}

Let us first focus on the case when $\phi = 0$, we have
\begin{equation}
  \epsilon_{\uparrow} = \epsilon_{\downarrow} = \Delta \cos\left(\frac{\epsilon_{\uparrow,\downarrow}L}{v}\right).
\end{equation}
Let us denote the common solution as $\epsilon$.

Now we consider $\phi\ll 1$ and denote $\delta \epsilon=\epsilon_{\uparrow} - \epsilon_{\downarrow}$.
By Eq.~(\ref{eq:up_spin}) and (\ref{eq:down_spin}) and condition (\ref{eq:condition}),
we have 
\begin{align}
  \delta \epsilon &= \Delta \left[\cos(\frac{\epsilon_\uparrow}{v}L-\frac{\phi}{2}) - 
  \cos(\frac{\epsilon_\downarrow L}{v}+\frac{\phi}{2})\right] \nonumber \\
  &= \Delta\cos\frac{\phi}{2}\left(\cos\frac{\epsilon_\uparrow L}{v} - \cos\frac{\epsilon_\downarrow L}{v} \right)
  + \Delta\sin\frac{\phi}{2}\left( \sin\frac{\epsilon_\uparrow L}{v} + \sin\frac{\epsilon_\downarrow L}{v} \right)
  \nonumber \\
  &\simeq  \Delta \sin(\frac{\epsilon L}{v}) \phi = \kappa v \phi, 
\end{align}
where
\begin{equation}
    \kappa = \frac{\sqrt{\Delta^2 - \epsilon^2}}{v}.
\end{equation}
This is valid up to first order in $\phi$.

In this situation, the operators $\hat{\sigma}^{+}$, $\hat{\sigma}^-$ and $\hat{\sigma}^z$ get simplified as
\begin{align}
  \hat{\sigma}^+ &=  2\exp\left[i\left(\kappa L - 1\right)\frac{\phi}{2} \right] 
    A_{\uparrow}^{*}A_{\downarrow}\gamma_{\uparrow}^{\dagger}\gamma_{\downarrow}  \\
  \hat{\sigma}^- &=  2\exp\left[-i\left(\kappa L - 1\right)\frac{\phi}{2} \right] 
    A_{\uparrow}A_{\downarrow}^*\gamma_{\uparrow}^{\dagger}\gamma_{\downarrow}  \\
  \hat{\sigma}^z &=
  \abs{A_{\uparrow}}^{2}(2\gamma_{\uparrow}^{\dagger}\gamma_{\uparrow}-1)-\abs{A_{\downarrow}}^{2}(2\gamma_{\downarrow}^{\dagger}\gamma_{\downarrow}-1).
\end{align}

Because of this, the total occupation number $N = \gamma_{\uparrow}^\dagger \gamma_\uparrow +
\gamma_{\downarrow}^\dagger \gamma_\downarrow$ becomes a good quantum number, namely $[N,H]=0$ where
$H = H_e + H_S$. The many body Hilbert space is spanned by the states $\ket{N\alpha}=\ket{N}\otimes\ket{\alpha}$
with $N=0,1,2$ labeling the occupation number of the Andreev bound state and $\alpha=+,-$ labeling
the eigenstates of $S_z$ of the impurity spin. 

The subspace for $N=0$ is spanned by $\ket{0+}$, and $\ket{0-}$. The Hamiltonian $H$ in this subspace is represented
by a $2$ by $2$ matrix
\begin{equation}
  H^{N=0} = -\epsilon - \frac{\abs{A_\uparrow}^2 - \abs{A_\downarrow}^2}{2}\left(J_{zz}\tau_z + J_{+z}\tau_{+} +
  J_{+z}^*\tau_- \right) 
\end{equation}
where $\tau_{\pm} = (\tau_x \pm  i \tau_y)/2$ with $\tau_{x,y,z}$ are Pauli matrices in this two dimensional subspaces. 

By using Eq.~(\ref{eq:A_updown}), we have
\begin{align}
  \abs{A_\uparrow}^2-\abs{A_{\downarrow}^2} &\simeq - 
  \frac{\epsilon}{2v(1+L\kappa)^2} \phi.
\end{align}

The subspace with $N=1$ is spanned by $\ket{\uparrow +}$, $\ket{\downarrow+}$, $\ket{\uparrow-}$, $\ket{\downarrow-}$. The Hamiltonian in this case
can be written as
\begin{equation}
  H^{N=1} = 2\exp\left[i\left(\kappa L 
  -1\right)\frac{\phi}{2}\right]A_{\uparrow}^{*}A_{\downarrow}(J_{z+}\tau_{z}+J_{++}\tau_{+}+J_{-+}\tau_{-})\rho_{+}+h.c.
\end{equation}
where $\rho_{\pm} = (\rho_x \pm i\rho_y)/2$ and $\rho_{x,y,z}$ are Pauli matrices in ${\ket{\uparrow},
\ket{\downarrow}}$ space, and we have neglected terms linear in $\phi\simeq 0$.

An unitary transformation $U=e^{-i\eta\rho_{z}/2}$ for some $\eta\in[0,\pi/2]$ can always be chosen
such that 
\begin{align}
  UH^{N=1}U^{\dagger}&=\frac{\kappa}{2(1+\kappa
  L)}\left[(J_{z+}\tau_{z}+J_{++}\tau_{+}+J_{-+}\tau_{-})\rho_z+h.c.\right] 
\end{align}
which has the same spectrum as $H^{N=1}$. The four eigenvalues are generically nondegenerate.

\end{widetext}


\begin{thebibliography}{18}
\bibitem{foot} Strictly speaking, the quasiparticle continuum also contributes to the Josephson effect. However, this contribution is necessarily $2\pi$ periodic so that any fractional Josephson effect originates from the subgap spectrum.
\end{thebibliography}
\end{document}